\documentclass[twocolumn,superscriptaddress,pre]{revtex4-1}
\usepackage{graphicx}
\usepackage{epstopdf}
\usepackage{amsmath}
\usepackage{subfigure}
\usepackage{amssymb}
\usepackage{mdframed}
\usepackage{color}
\usepackage[usenames,dvipsnames,table]{xcolor}
\makeatletter

\newcommand{\Rmnum}[1]{\expandafter\@slowromancap\romannumeral #1@}

\makeatother
\makeatletter
\renewcommand{\@biblabel}[1]{\quad#1.}
\makeatother

\makeatletter
\renewcommand{\@biblabel}[1]{\quad#1.}
\makeatother

\begin{document}

\title{ \textbf{The arrival of the frequent: how bias in genotype-phenotype maps can steer populations to local optima}
}
\author{Steffen Schaper} 
\affiliation{ Rudolf Peierls Centre for Theoretical Physics, University of Oxford, Oxford, UK}
\affiliation{
Aachen Institute for Advanced Study in Computational Engineering Science (AICES), RWTH Aachen University, Aachen, Germany}
\author{Ard A. Louis}
\affiliation{Rudolf Peierls Centre for Theoretical Physics, University of Oxford, Oxford, UK}
\email{E-mail: ard.louis@physics.ox.ac.uk}
\date{\today}
\begin{abstract} Genotype-phenotype (GP) maps specify how the random mutations that change genotypes generate
variation by altering phenotypes, which, in turn, can trigger selection. Many GP maps share the following general
properties: 1) The number of genotypes $N_G$ is much larger than the number of selectable phenotypes; 2) Neutral
exploration changes the variation that is accessible to the population; 3) The distribution of phenotype frequencies
$F_p=N_p/N_G$, with $N_p$ the number of genotypes mapping onto phenotype $p$, is highly biased: the majority of
genotypes map to only a small minority of the phenotypes. Here we explore how these properties
affect the evolutionary dynamics of haploid Wright-Fisher models
that are coupled to a simplified and general random GP map or to a more complex RNA sequence to secondary structure map.  For both maps the probability of a mutation leading to  a phenotype $p$ scales to first order as $F_p$, although for the RNA map there are further correlations as well.  By using mean-field theory, supported by computer simulations, we show that the discovery time $T_p$ of a phenotype $p$ similarly scales to first order as $1/F_p$ for a wide range of population sizes and mutation rates in both the monomorphic and polymorphic regimes.  These differences in the rate at which variation arises can vary over many orders of magnitude. Phenotypic variation with a larger $F_p$ is therefore be much more likely to arise than variation with a small $F_p$. We show, using the RNA model, that  frequent phenotypes (with larger $F_p$) can fix in a population 
even when alternative, but less frequent, phenotypes with much higher fitness are potentially accessible. In other words,  if the fittest never `arrive' on the timescales of evolutionary change, then they can't fix.  We call this highly non-ergodic effect the `arrival of the frequent'.
\end{abstract}

%
\maketitle

\begin{figure*}[!ht] 
\begin{center} 
\includegraphics{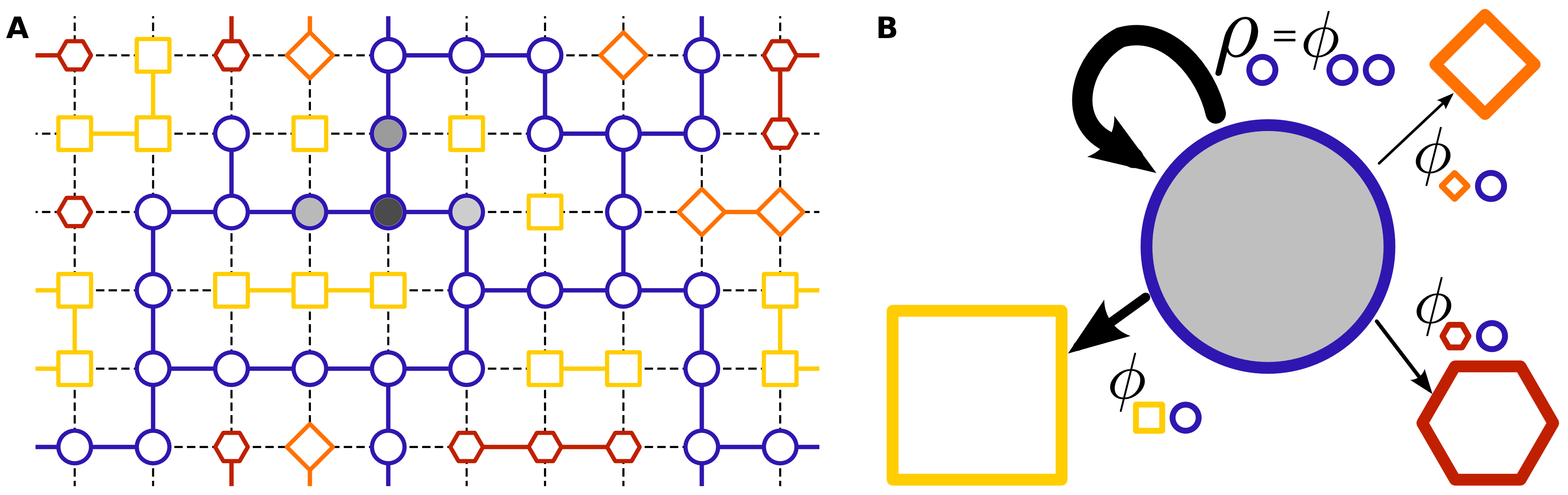} 
\end{center} 
\caption{{\bf Illustration of the mean field approximation.} \textit{A)} An example genotype space: Each point
corresponds to a unique genotype; shape and color of the marker indicate the phenotype. Genotypes joined by edges can be
interconverted by single mutations. Edges for neutral mutations share the color of the (conserved) phenotype,
non-neutral mutations are shown as black dashed lines. The shading of the genotypes illustrates the number of
individuals carrying the respective genotype in a hypothetical population. The mutations away from the genotypes
occupied by the population determine the accessible phenotypes. \textit{B)} Our meanfield approximation averages over the
internal structure of neutral spaces. So neutral spaces are represented by the markers of their phenotypes only, with
the size representing the neutral space size (ie. number of genotypes in the space). The uniform shading of the blue
neutral space implies that in the meanfield approximation, the population is assumed to continually explore the
neighbourhood of its entire neutral space. Mutational outcomes are thus determined from the local frequencies of
phenotypes around the neutral space, as measured by the $\phi_{pq}$ coefficients. This mean field approximation allows us to derive analytic forms that can be compared to simulations of the full GP map.} 
\label{fig:meanfield_illustration} 
\end{figure*}
\section*{Introduction} \par Darwin's account of biological evolution \cite{darwin1859origin} stressed the importance of
natural selection: If some individuals are better adapted to their environment than their competitors, their offspring
will come to dominate the population. The fittest survive and the less fit go extinct. Yet selection alone is not
sufficient to drive evolution because natural selection reduces the very variation that it requires to operate. It was
only recognised well after Darwin's day~\cite{devries1904mutation}, in part through the success of the Modern Synthesis,
that the fuel for selection is provided by mutations that make offspring genetically different from their parents.
Crucially, mutations change genetically stored information (the \emph{genotype}) while selection operates on the
physical expression of this information (the \emph{phenotype}). Understanding the relation between genotypes and
phenotypes -- the GP map -- is therefore crucial to understanding evolutionary dynamics \cite{alberch_genes_1991}.

\par GP mappings have been studied at different levels of abstraction~\cite{wagner2005book} The most basic systems are
concerned with the sequence-to-structure(-to-function) relation of single molecules such as RNA
\cite{schuster1994neutralnetworks} or proteins \cite{li1996hpmodel,england2003structural,ferrada2010proteinfunction},
but higher-level systems such as protein complexes \cite{ahnert2010selfassembly}, gene-regulatory
networks~\cite{raman2011signalling} and developmental networks \cite{borenstein2008bias} have also been studied. Even
though these GP maps arise in quite different contexts, they share several interesting properties:

\par 1) Most basically, the number of possible genotypes $N_G$ is typically much greater than the number of possible
phenotypes $N_P$, so the map is many-to-one. As a consequence, many mutations may conserve the phenotype, leading to
mutational robustness. Important prior work has linked such robustness to the concept of neutral spaces, namely the set
of all genotypes that map to a particular phenotype, with the additional property that they be linked by neutral
mutations~\cite{wagner2005book,schuster1994neutralnetworks,ferrada2012comparison}.

\par 2) Even though $N_P \ll N_G$, the accessible genetic neighbourhood of a single genotype $g$ that generates a given
phenotype $p$ may include significantly fewer alternative phenotypes (potential variation) than is found in the
neighbourhood of the (neutral) set ${\cal N}_p$ of all $N_p=|{\cal N}_p|$ genotypes that map onto phenotype $p$.
Exploration of a neutral space can therefore increase the variety of phenotypes discovered by a
population~\cite{wagner2008paradox,wagner2008neutralism}.

\par 3) Perhaps the most striking commonality of these GP maps is a strong bias in assignment of genotypes to
phenotypes: Most phenotypes are realised by a tiny proportion of all genotypes, while most genotypes map into a small
fraction of all phenotypes. This property is shared by all the GP maps we noted before. Typically the number $N_p$ of
genotypes per phenotype $p$ and the related phenotype frequencies $F_p = N_p/N_G$ can vary over many orders of
magnitude. Such huge variations are likely to have an effect on the course of evolution.

\par In this paper we study the evolutionary dynamics of a classical Wright-Fisher model, but with explicit microscopic
GP maps that capture the three generic properties of such maps introduced above. Motivated by the strong bias in the
distribution of the $F_p$ observed for many GP maps, we derive a mean-field like approximation for the average
probability $\phi_{pq}$ that a mutation will change a genotype that generates phenotype $q$ into one that generates
phenotype $p$. This approximation greatly simplifies the dynamics, allowing us to calculate analytic expressions for
quantities such as the median time $T_p$ for phenotype $p$ to first appear in the population as a function of population
size $N$, the point mutation rate $\mu$, genome length $L$ and the mutation probabilities $\phi_{pq}$.

\par These approximations are then tested against extensive simulations of two models: firstly, a simple GP map where
the genotypes are randomly assigned to phenotypes according to a pre-determined distribution for the frequencies $F_p$
and secondly, the well-known mapping of RNA sequence to secondary
structure~\cite{schuster1994neutralnetworks,ViennaPackage,wagner2005book}, which is more complex, but also more
biologically realistic. We focus on the case where a population of $N$ individuals has initially equilibrated at a
fitness maximum given by phenotype $q$, and then measure the median time $T_p$ for alternative phenotype $p$ to first
arise in the population.

\par Our analytic expressions agree quantitatively with the simulations in the polymorphic limit where $N L \mu \gg 1$,
and also in the opposite monomorphic limit $N L \mu \ll 1$. In between these regimes a single scaling factor must be
included. In all regimes the median discovery time $T_p \propto 1/\phi_{pq}$. For the random model $\phi_{pq} \approx
F_p$; this scaling also holds for the more complex RNA mapping, although there is significantly more scatter due to
local correlations within the neutral spaces and for some phenotypes we find $\phi_{pq}=0$ even though $F_p$ is large (this can be due to biophysical 
constraints explained for example in ref.~\cite{schaper2011contingency}.
Despite such higher order effects, the variation of the $F_p$ over many orders translates directly into the $T_p$. More
frequent (higher $F_p$) phenotypes are therefore discovered more rapidly and more often along evolutionary trajectories.
In this way the structure of the GP map can play a key role in determining evolutionary outcomes.

\par Finally, we employ the RNA GP map to study the case where two phenotypes $p1$ and $p2$ are both more fit than the
source phenotype $q$, but where $F_{p1} \gg F_{p2}$ (or more accurately $\phi_{p1q} \gg \phi_{p2q}$). Direct simulations show that phenotype $p1$, which is more
frequent, is much more likely to fix in the population, even if its fitness is much lower than that of $p2$, an effect
we call `the arrival of the frequent'.

\begin{figure*}[t] 
\begin{center} 
\includegraphics[width=17cm]{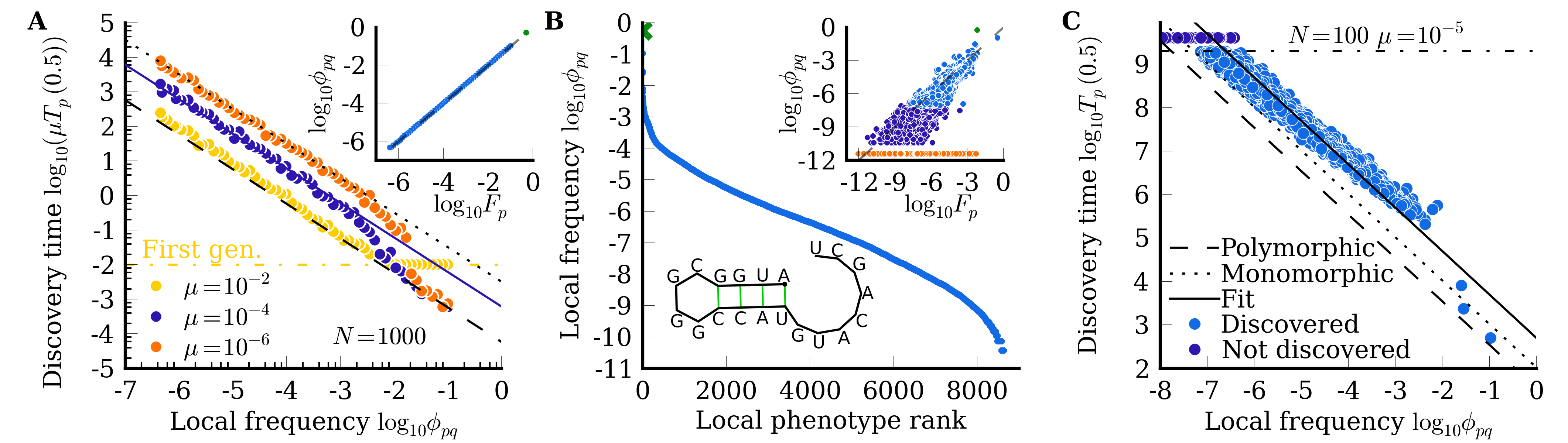} 
\end{center} 
\caption{{\bf Test of the meanfield model.} \textit{A)} Median discovery times $T_p$ for the random GP map averaged over
100 simulations with $N=1000$ and varying mutation rates. Note that the y-axis is scaled with $\mu$. In the the
polymorphic limit ($\mu=10^{-2}$), Eq.~\protect\eqref{eq_Tp} (dashed line) describes discovery times well for $\phi_{pq}
< 1/(NL\mu)$. Phenotypes with larger $\phi_{pq}$ are part of the standing variation typically found in the first
generation (yellow dash-dotted line). In the monomorphic limit ($\mu=10^{-6}$),
Eq.~\protect\eqref{eq:monomorphic_discovery_time} (dotted line) quantitatively describes $T_p$ for $\phi_{pq} \ll
\phi_L$, whereas Eq.~\protect\eqref{eq_Tp} tracks the simulation data with just one fit parameter $\gamma = 0.099$
multiplying $N$ for the intermediate regime with $\mu=10^{-4}$ (solid line). For $\phi_{pq} \gtrsim \phi_L$ the curves
follow Eq.~\protect\eqref{eq_Tp}, for reasons described in the text. \textit{Inset}: For the random GP map the local
phenotype frequency $\phi_{pq}$ correlates very well with the global frequency $F_p$. \textit{B)} Local frequency
$\phi_{pq}$ ranked for the $8639$ phenotypes that link with single point mutations from the $|{\cal{N}}_q| =
460,557,583$ genotypes that map to this RNA structure; an example sequence from ${\cal{N}}_q$ is shown in the figure. \textit{Inset}: The
local connections $\phi_{pq}$ are roughly proportional to the global frequency $F_p$, but there is significant scatter
due to the internal correlations of the RNA neutral spaces. Organge points depict the $2580$ phenotypes for which
$\phi_{pq}=0$. Light blue points depict the $4933$ phenotypes that are discovered in our simulations, and the dark blue
points depict the $3705$ accessible phenotypes that are not found ($q$ itself is shown in green). \textit{C)} Simulations
of $T_p$ (blue dots) versus $\phi_{pq}$ for the RNA phenotype shown in B), compared to Eq.~\protect\eqref{eq_Tp} (solid
line) with a factor $\gamma = 0.070$ multiplying $N$. Here $N=100$, $\mu = 10^{-5}$ and the simulations were run for
$2\times 10^9$ generations. Also shown are the purely polymorphic (dashed) and monomorphic (dotted) predictions. 
Dark blue dots above $2\times 10^9$ (dot-dashed line) depict some of the $3705$ accessible phenotypes that are not found (as
can be seen in see the inset of B). We estimate that about $10^{13}$ generations would be needed to find the phenotypes
with the smallest $\phi_{pq} \ne 0$. } 
\label{fig:mediantime_nullmodel}
\end{figure*}

\section*{Results}

\subsection*{Theoretical framework}

\par We study the evolution of a population of $N$ asexual haploid individuals. Each individual $i$ carries a genotype
$g_i$ of $L$ letters taken from an alphabet of size $K$. The individual's phenotype $p_i$ is determined from $g_i$ via
the GP map. The population evolves in in discrete, non-overlapping generations according to the classical Wright-Fisher
model for haploid individuals: At each generation $T$, $N$ parents are drawn with replacement with probability
proportional to their fitness $1 + s_i$ with the constraint that the population size (or carrying capacity) $N$ is
fixed. Each parent gives rise to one offspring, and the offspring make up the population for the next generation. During
reproduction, each base in the genotype of length $L$ mutates to a random alternative base with probability $\mu$. The
number of mutations (that is, the Hamming distance) $d$ between parent and offspring is thus distributed binomially
according to $h(d) = {L \choose d} \mu^d (1-\mu)^{L-d}$. In this way the set $\{g_i\}$ of $N$ genotypes changes at each
generation.

\par The expected number of individuals with phenotype $p$ that arises at generation $t$ can be written as:

\begin{equation} 
\label{eq1} 
m_p(t) = \sum_i^N \sum_{d=1}^L h(d) \Phi_p(g_i,s_i,d)
\end{equation} 
where $\Phi_p(g_i,s_i,d)$ is the probability that a $d$ fold mutation of genotype $g_i$ (selected for reproduction
according to fitness $1 + s_i$) generates an individual with phenotype $p$. It takes into account the mutational
connections between the $N_G = K^L$ genotypes that make up the GP map. The probability of not finding $p$ is
approximately given by the Poisson distribution as $\exp(-m_p(t))$.

\par While exact, these dynamic expressions depend implicitly on time through stochastic changes in the set $\{g_i\}$,
and are typically very hard to solve. In order to gain intuitive insight, we employ a number of simplifications and
approximations, motivated in part by the general properties of GP maps discussed in the introduction. First, we assume
that $L\mu \ll 1$  so that for $d>1$, $h(d) \ll h(1) \approx L \mu$, which means that we can ignore higher order mutations (terms with $d>1$ in Eq.~(\ref{eq1})). For a given source phenotype $q$ (where the fitnesses of all genotypes mapping into $q$ are
equal, and so we take $1 +s_q=1$ for simplicity) we can then calculate the mean probability $\phi_{pq}$ that a single
point mutation will generate another phenotype $p$: 

\begin{equation} 
	\phi_{pq} = \frac{1}{N_q} \sum_{i=1}^{N_q} \Phi_p(g_i,0,1) 
\end{equation} 
where the sum is over the set ${\cal N}_q$ of all $N_q$ genotypes that generate phenotype $q$ (see also
Fig.~\ref{fig:meanfield_illustration}). It is convenient to introduce the robustness of phenotype $q$ as the average
probability over all ${\cal N}_q$ of neutral mutations: $\rho = \phi_{qq}$. If we consider the case where at generation
$t-1$ the whole population is on ${\cal N}_q$, then Eq.~\eqref{eq1} simplifies in this mean-field (or pre-averaged) approximation to: 

\begin{equation}
\label{eq_mp} 
	m_p(t) = N L \mu \phi_{pq}
\end{equation}

\subsubsection*{The polymorphic limit} 

\par If $N L \mu \gg 1$ then the population naturally spreads over different genotypes, a regime called the polymorphic
limit. Consider the case where $1 + s_p = \delta_{qp}$ so that the population remains on ${\cal N}_q$, which is one way
to model neutral exploration. In the mean-field approximation the expected number of individuals with phenotype $p$
produced per generation is now independent of time, and given by Eq.\eqref{eq_mp}, as long as double mutations can be
ignored. The time $T_p(\alpha)$ when on average the probability of having discovered $p$ is $\alpha$ (so that the median
discovery time of $p$ is $T_p(1/2)$) is then given by:

\begin{equation}
	\label{eq_Tp} T_p(\alpha) = \frac{-\log(1-\alpha)}{N L \mu \phi_{pq}}
\end{equation}

Eqns.~(\ref{eq_mp} - \ref{eq_Tp}) should provide a good approximation of the full dynamics in the limit that $N$ is
large enough that variations between individual genotypes $g_i \in {\cal N}_q$ are averaged out, in other words, for the
case where the 1-mutant neighbourhood of the population is similar to that of the whole neutral space.

\subsubsection*{The monomorphic limit} 

\par Neutral spaces can be astronomically large~\cite{joerg2008nnsize}, much bigger than even the largest viral or
bacterial populations. In that case, the local neighborhood of the population may not be fully representative of the neighborhood of the
entire space. This scenario can most easily be understood in the monomorphic limit where mutants are rare, $NL\mu \ll
1$, and exploration is dominated by genetic drift. Every neutral mutant has a probability of $1/N$ to go to fixation,
allowing the population to move to a new genotype. Thus the timescale of fixations is Kimura's famous result
\cite{kimura1985book} $\tau_f = 1/(L\mu\rho)$, where the robustness $\rho$ is the probability that a mutation is
neutral, so that $L\mu\rho$ is the rate of neutral mutations.

\par Between fixations, the population undergoes periods of genotypic stasis in which only the 1-mutant neighborhood of
the current genotype $g$ is explored by (rare) mutations. As there are $(K-1)L$ adjacent genotypes, the timescale of this
exploration is $\tau_e = (K-1)L/(NL\mu) = (K-1)/(N\mu)$. 

\par It is instructive to compare the ratio $\xi$ of these two time-scales, defined via 
\begin{equation} 
	\xi = \frac{\tau_f}{\tau_e} = \frac{N}{(K-1)L\rho} \approx \frac{N}{L} 
\end{equation} 
We can use this dimensionless ratio to distinguish between different dynamic regimes. If $\xi \gg 1$, fixation takes
much longer than exploration. If we define $n_p^g$ as the number of local neighbours of the genotype $g$ mapping to
phenotype $p$ for the current population, then in this limit, phenotypes with $n_p^g > 0$ are produced continuously (on
a time-scale given by $\tau_e$) until the population moves to a different genotype. The dynamics under strong genetic
drift therefore induce short-term correlations in the mutant phenotypes. Since $\xi \approx N/L$, we call this regime
the large population limit.

\par In the opposite extreme $\xi \ll 1$, which we call the large genome limit, the population typically moves to a
different genotype before all accessible mutants have been explored. In this regime, we do not expect short-term
correlations in the mutant phenotypes, simply because every mutant occurs only very rarely.

\par Actual discovery and neutral fixation times can show strong fluctuations. As our evolutionary process is a Markov
process -- the next set of mutants depends only on the parents, not on earlier mutants -- the first discovery time of a
neighbour genotype as well as the arrival time of the neutral mutant ``destined'' to be fixed, are distributed
geometrically (or exponentially in a model with continuous time). Thus the mean of $\tau_e$ or $\tau_f$ is equal to the
respective standard deviation, and any particular evolutionary trajectory can be very different from the average
behaviour.

\par Let $\tau$ be the actual time the population stays at the current genotype. In the continuous time approximation,
$\tau$ is distributed exponentially with mean $\tau_f$. If the genotype $g$ has $n_p^g$ mutations leading to $p$ then the
probability that  $p$ is found during this time is $1-\exp(-n_p^g\tau/\tau_e)$. Integrating over the distribution of
$\tau$, we have the probability $P(n_p^g)$ that phenotype $p$ is discovered before the next neutral fixation:
\begin{equation} 
P(n_p^g) = \int_0^\infty \frac{d\tau}{\tau_f} (1-e^{-n_p^g\tau/\tau_e}) e^{-\tau/\tau_f} = 1-\frac{1}{1+n_p^g \xi} 
\label{eq:monomorphic_discovery_probability} 
\end{equation}

\par If fixations are the rate-limiting step (ie. $\xi \gg 1$), $P \rightarrow 1$ if $n_p^g \neq 0$, as each
neighborhood is searched exhaustively before the population moves on. On the other hand, if fixation is faster than
exploration ($\xi \ll 1$), the introduction of alternative phenotypes is determined by random fluctuations, as most available
mutants are not produced. To leading order, we find $P(n_p^g) \approx n_p^g\xi = Nn_p^g/((K-1)L\rho)$. We note that the
inverse dependence on $\rho$ arises from $\tau_f$: More robust neutral spaces are explored faster, but therefore less
thoroughly.

\par The dynamics in the monomorphic regime are thus relatively straightforward. But whether some new phenotype $p$ is
discovered still depends on the structure of the neutral space which in turn determines how the available phenotypes change
upon a neutral fixation. To describe this structure, we turn again to a mean-field approximation: The mutational
neighborhood of each particular genotype $g \in \mathcal{N}_q$ resembles the average over $\mathcal{N}_q$. As the mean
number of mutations per genotype leading to $p$ is given by $\bar{n}_{pq} = (K-1)L\phi_{pq}$, the probability that $p$
is accessible after a neutral fixation is $1-\exp(-\bar{n}_{pq}) \approx \bar{n}_{pq}$ (the approximation is valid
provided $n_{pq} \ll 1$, that is $p$ is not accessible from every genotype in the source neutral space; of course, this
is just the condition we are interested in, as otherwise neutral exploration would not typically be necessary for phenotype $p$ to arise).

\par Over a large number of generations ($\tau \gg \tau_f$), a monomorphic population explores its neutral space
uniformly~\cite{vanNimwegen1999robustness}. Assuming that $n_p^g >1$ can be ignored in practice, we have $T_p(\alpha) =
-\tau_f \log(1-\alpha)/(n_{pq} P(1))$. The first discovery time in the large population limit becomes: 

\begin{equation} 
	T_p(\alpha) = \frac{-\tau_f \log(1-\alpha)}{ n_{pq}} = \frac{-\log(1-\alpha)}{L^2(K-1) \mu \rho \phi_{pq}}
\label{eq:monomorphic_discovery_time} 
\end{equation} 
whereas in the large genome limit we obtain

\begin{equation}
\label{eq:Tpmono2} 
	T_p(\alpha) = \frac{-\tau_e \log(1-\alpha)}{ n_{pq}} = \frac{-\log(1-\alpha)}{N L \mu \phi_{pq} }
\end{equation} 
which has the same form as the polymorphic limit, Eq.~\eqref{eq_Tp}: When the population is too small (compared to the
genome length), the exploration of each genotype's mutational neighborhood is typically incomplete. Then, just as in the
polymorphic limit, only random fluctuations determine which accessible genotypes are actually realized by the population.

\par Finally, let us compare our results for large populations in the monomorphic and polymorphic limits. Most
importantly, in both cases $T_p$ is inversely proportional to $\phi_{pq}$: Rare phenotypes are hard to find. Comparing
Equations \eqref{eq_Tp} and \eqref{eq:monomorphic_discovery_time}, the only difference is that $N$ in the polymorphic
regime is replaced by $L(K-1)\rho$ in the monomorphic limit. This difference is intuitive: When the population is
diverse, every new individual helps exploration and reduces discovery times. But if all individuals have the same
genotype, simply having ``more of the same'' does not make neutral exploration faster. However, repeated mutants may
influence the fixation of adaptive phenotypes.

\par These results suggest that for intermediate $NL\mu$ there should be a smooth transition between these two regimes.
To quantify the crossover we introduce a factor $\gamma$ that multiplies $N$ in Eq.\eqref{eq_Tp}; we expect that $\gamma
\rightarrow 1$ as either $NL\mu$ becomes very large (the polymorphic limit) or $N \ll L$ (the large genome limit), and
that $\gamma \rightarrow (K-1)L\rho/N$ as $NL\mu \ll 1$ and $N \gg L$ (the large population monomorphic limit).

\subsection*{Simulations in model GP maps}

In order to test our mean-field theory we study two kinds of GP maps that both include the generic properties of GP maps
that we introduced earlier.

\subsubsection*{Random GP map} 

\par In the random GP map, the total number of phenotypes $N_P$ and the frequencies $\{F_p\}$ can be set arbitrarily
(subject to the normalization constraint $\sum_{p=1}^{N_P} F_p = 1$). The $K^L \times F_p$ genotypes mapping into
phenotype $p$ are distributed randomly in genotype space. The statistical properties of the map are thus determined by
the parameters $L$, $K$, and the set $\{F_p\}$.

\par Studying this map has two motivations: First, ignoring some biophysical detail may help illuminate generic features
shared by the systems described in the introduction. Second, a simple model may clarify which deviations from our theory
arise from population dynamic effects rather than from detailed (and system-specific) structure in the GP map.

\par In this simple model, correlations between genotypes are absent, facilitating analysis of the resulting neutral
spaces. For example, $\phi_{pq} = F_p$ is a good approximation as long as $N_P \ll N_G$ and $N_q, N_p \gg 1$. Also,
there is a percolation threshold $\lambda(K) = 1 - K^{-1/(K-1)}$: thus only phenotypes with $F_q > \lambda(K)$ have
completely connected neutral spaces \cite{reidys1997percolation}.

\par Here we study a particular random GP map with $L=12$, and $K=4$ (as in DNA and RNA) so that there are
$N_G=4^{12}\approx1.68 \times 10^7$ genotypes. These map onto $N_P=58$ phenotypes distributed with frequencies $F_p \propto
1.2^{- p}$. The $F_p$ vary over about $5$ orders of magnitude, a range similar to the $F_p$ of $L=12$ RNA (see also Fig.
\ref{fig:si_nullmodel_statics}). To make sure that the largest neutral space percolates, its frequency is set separately
as $F_1=0.5 > \lambda(4) = 0.37$. For several values of $\mu$, we simulated $N=1000$ individuals for up to $7\times
10^{10}$ generations. The fitness was set as $1 + s_p=\delta_{p,1}$ so that we are effectively modelling neutral
exploration on the space ${\cal N}_1$, which is convenient for measuring all $T_p$. We measured first discovery times
for the $57$ alternative phenotypes over $100$ independent simulations to obtain the median time $T_p$.

\par Figure~\ref{fig:mediantime_nullmodel}A depicts these median discovery times $T_p$ for simulations ranging from the
polymorphic regime $N L \mu \gg 1$ to the monomorphic limit $N L \mu \ll 1$. We note the following:

\par 1) For all regimes the $T_p$ vary over many orders of magnitude, but they are found in fewer generations for larger
$\mu$.

\par 2) Locally frequent phenotypes (i.e.~those with high $\phi_{pq}$) are much easier to discover. The inset of
Figure~\ref{fig:mediantime_nullmodel}A shows that $\phi_{pq} \approx F_p$, so this conclusion carries over to frequent
phenotypes with large $F_p$.

\par 3) A subset of the phenotypes with $\phi_{pq} > \phi_L \equiv 1/(K-1) L \approx 0.028$ are likely to be in the
one-mutation neighbourhood of any genotype. In the monomorphic regime these are are then found by exploration of a
genome so that $T_p$ is given by Eq.~\eqref{eq:Tpmono2}, which has the same form as the polymorphic limit,
Eq.~\eqref{eq_Tp}, as can be seen in Fig.~\ref{fig:mediantime_nullmodel}A. Discovery times cross over to the regime
where neutral exploration is required when $\phi_{pq} \ll \phi_L$. Such behaviour can be viewed as a finite size effect:
$N_P$ typically increases with $L$. Therefore the largest $F_p$ will likely
decrease for larger systems, so that a smaller fraction of phenotypes can be found without neutral exploration.

\par 4) In the fully polymorphic regime where each individual essentially explores independently, any phenotype with
$\phi_{pq} > 1/(NL\mu)$ is likely to be part of immediately accessible {\em standing
variation}~\cite{barrett2008adaptation} in the initial population, and is therefore found quickly. Indeed, in
Figure~\ref{fig:mediantime_nullmodel}A for $\mu=10^{-2}$, where $NL\mu = 120$, these phenotypes are typically found in
one or two generations on average. However, for rarer phenotypes, where neutral exploration is important, the $T_p$ are
well approximated by Eq.~\eqref{eq_Tp}. Again, the fraction of phenotypes that are immediately accessible should
decrease for larger $L$.

\par 5) In the intermediate regime $\mu = 10^{-4}$, where $N L \mu = 1.2$, the population spreads over more phenotypes
than in the monomorphic regime, but over fewer than in the polymorphic regime. Thus the crossover to the regime where neutral
exploration is important occurs at a smaller $\phi_{pq}$ than for the monomorphic regime. In this intermediate $\mu$
regime neither Eq.~\eqref{eq_Tp} nor Eq.~\eqref{eq:monomorphic_discovery_time} suffices. Instead, we use the previously
introduced factor $\gamma$ that multiplies $N$ in Eq.~\eqref{eq_Tp} to achieve quantitative accuracy. In the
supporting information 
we explore the scaling of $\gamma$ with the parameters $N, L,
\mu$, and also study the $T_p$ in the large genome limit $\xi \ll 1$, showing that Eq.~\eqref{eq:Tpmono2} holds, as
derived in the previous section.

\par In summary then, our theory derived in the previous section accurately describes the median discovery time $T_p$ of
this simple random GP map as a function of the parameters $N, \mu, \phi_{pq}$. We find that $\phi_{pq} \approx
F_p$, and thus $T_p \sim 1/F_p$ in all regimes studied. The more frequent the phenotype, the earlier (and more often,
see Fig.~\ref{fig:si_mutant_count}) it appears as potentially selectable variation in an evolving population. Given the
success of our theory for the random model, we now will test our theory and conclusions for a more complex GP map.

\begin{figure*}[t] 
\begin{center} 
\includegraphics{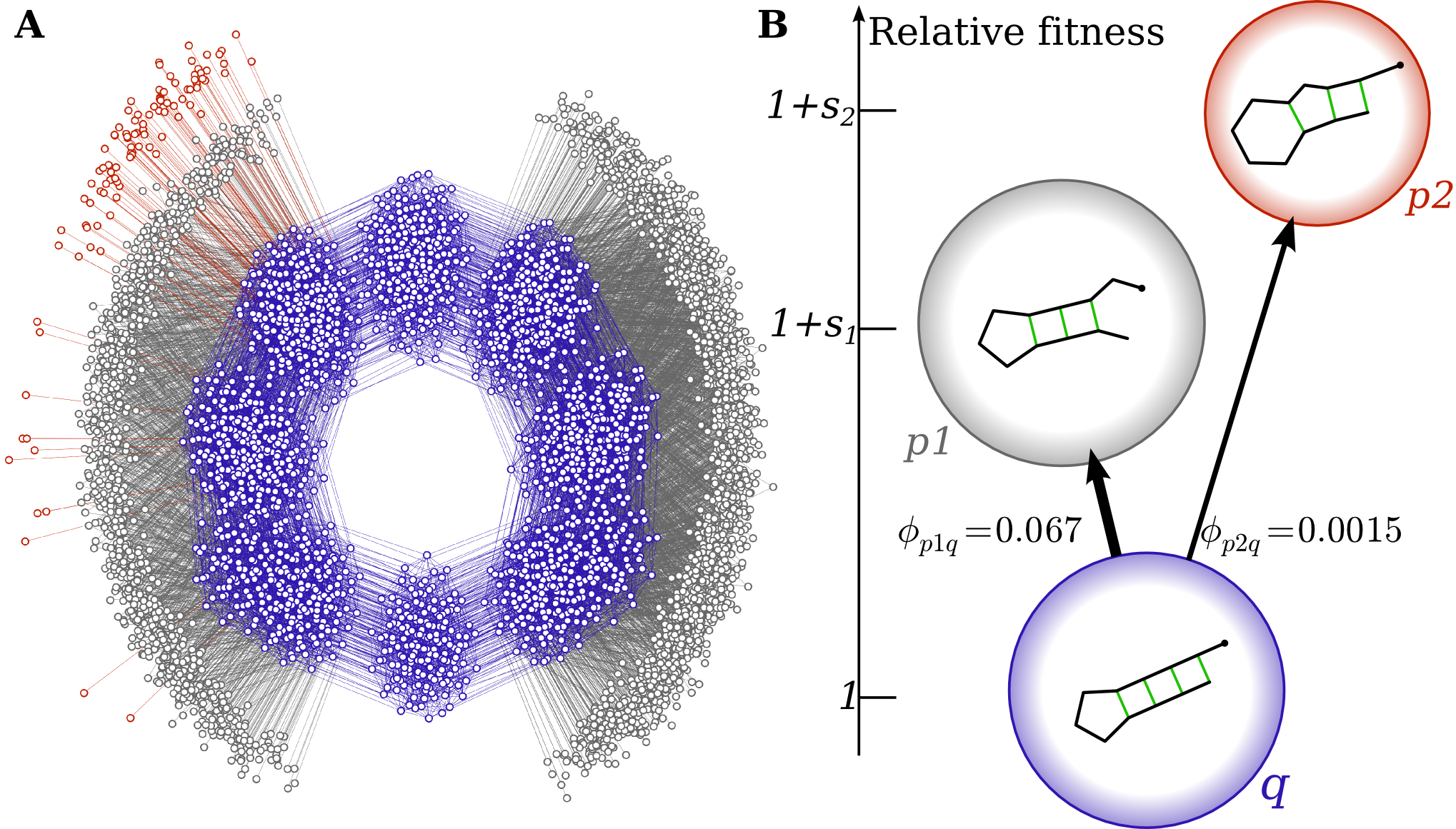} 
\end{center} 
\caption{{\bf Interconnections of neutral spaces in RNA influence evolutionary trajectories.}  \textit{A}) $L=12$ RNA 
neutral component for phenotype $q$ with $N_q=1932$ genotypes (drawn in blue). Lines depict single mutations to itself, 
or to two alternative phenotypes $p1$ (grey) and $p2$ (red). The genotypes were ordered using the Fruchterman-Reingold algorithm~\protect\cite{fruchterman-reingold1991algorithm}. \textit{B}) Illustration of the fitness landscape.} 
\label{fig:RNA} 
\end{figure*}

\subsubsection*{RNA secondary structure mapping} 

\par One of the best studied GP mappings has RNA genotypes of length $L$ made up of nucleotides G, C, U and A. The
phenotypes are the minimum free-energy secondary structures for the sequences, which can be efficiently
calculated~\cite{ViennaPackage}. The number of genotypes grows as $4^L$, while the number of phenotypes is thought to
grow roughly as $N_P \sim 1.8^L$~\cite{wagner2005book} so that $N_P \ll N_G$. Moreover, sampling and exact
enumerations\cite{schuster1994neutralnetworks,cowperthwaite2008ascent,schaper2011contingency} have shown that the
distribution of phenotype frequencies $F_p$ is highly biased, with a small fraction of phenotypes taking up the majority
of genotypes. The neutral spaces ${\cal N}_q$ are typically broken up into a number of large components that are
connected by single point mutations that allow neutral
exploration~\cite{schaper2011contingency,cowperthwaite2008ascent}. By exhaustive enumeration of the $L=20$ RNA mapping
(see also Fig.~\ref{fig:si_rna_static}) we calculate the $\phi_{pq}$ between several neutral components of the $11,219$
distinct secondary structures that the $N_G=4^{20} \approx 1.1 \times 10^{12}$ genotypes map to.

\par Figure~\ref{fig:mediantime_nullmodel}b shows the $\phi_{pq}$ for the largest component of the phenotype $q$ drawn
in the figure. This phenotype is ranked as the 3rd most frequent for $L=20$ and exhibits behaviour typical of this
system. First, the $\phi_{pq}$ vary over many orders of magnitude. Second, as shown in the inset if $\phi_{pq} \neq 0$,
then the \emph{local} $\phi_{pq}$ are, to first order, proportional to the \emph{global} $F_p$. Finally, this neutral
space connects to just over $75\%$ of the total $N_P=11,219$ phenotypes in this particular map: Some $\phi_{pq}$ are
zero even though $F_p$ can be quite large. Generally, the number of phenotypes that can be reached from ${\cal N}_q$
increases with $F_q$~\cite{schaper2011contingency,wagner2008paradox}.

\par We performed extensive simulations of the $L=20$ RNA system. Typical results are shown in
Figure~\ref{fig:mediantime_nullmodel}B. {\it First}, we note that the median discovery times vary over many orders of
magnitude. The most frequent are found in a median time of  $T_p \approx 10^{3}$ generations while after the maximum measured time of $2 \times 10^{9}$ generations, over $42 \%$ of the directly accessible phenotypes (with $\phi_{pq} \neq 0$) have still not been found. We estimate that over $10^{13}$ generations would be needed to discover all accessible phenotypes, giving a ten order of magnitude range in the $T_p$.  {\it Second}, the
local frequency $\phi_{pq}$ is a good predictor for ranking $T_p$; further, the criterion $\phi_{pq}=0$  accurately predicts which
phenotypes are \emph{not} discovered (see also Fig. \ref{fig:si_rna_TvsF}). However, in contrast to the random GP map,
the $T_p$ are discovered at a slower rate than predicted by Eq.~\protect\eqref{eq:monomorphic_discovery_time}. Instead,
we use a single $\gamma < 3 L\rho/N$ to renormalise $N$ in Eq.~\protect\eqref{eq_Tp}. This slower discovery rate
reflects the internal structure of the RNA: similar genotypes typically have similar mutational neighbourhoods
\cite{huynen1996exploration}, and so the population needs to neutrally explore longer in order to find novelty.
Nevertheless, a single $\gamma$ factor yields a remarkably good fit for all the different phenotypes $p$ (something we
find for all source phenotypes $q$ we have so far studied). Finally, we note that the three most frequent phenotypes are
found relatively faster because they satisfy $\phi_{pq} \gtrsim \phi_L$. As expected, for this larger system the
fraction of phenotypes for which this holds is lower than for the random GP map with smaller $L$.

\par Overall, the evolutionary dynamics of this rather complex RNA system resembles that of the much simpler random GP
map. Most importantly, the discovery times vary over many orders of magnitude. More precisely, as long as $\phi_{pq}
\neq 0$, $T_p \propto 1/\phi_{pq}$ for both the monomorphic and polymorphic regimes: Phenotypic bias leads to a simple,
systematic ordering in the discovery of novel phenotypes.

\begin{figure*}[t] 
\begin{center} 
\includegraphics{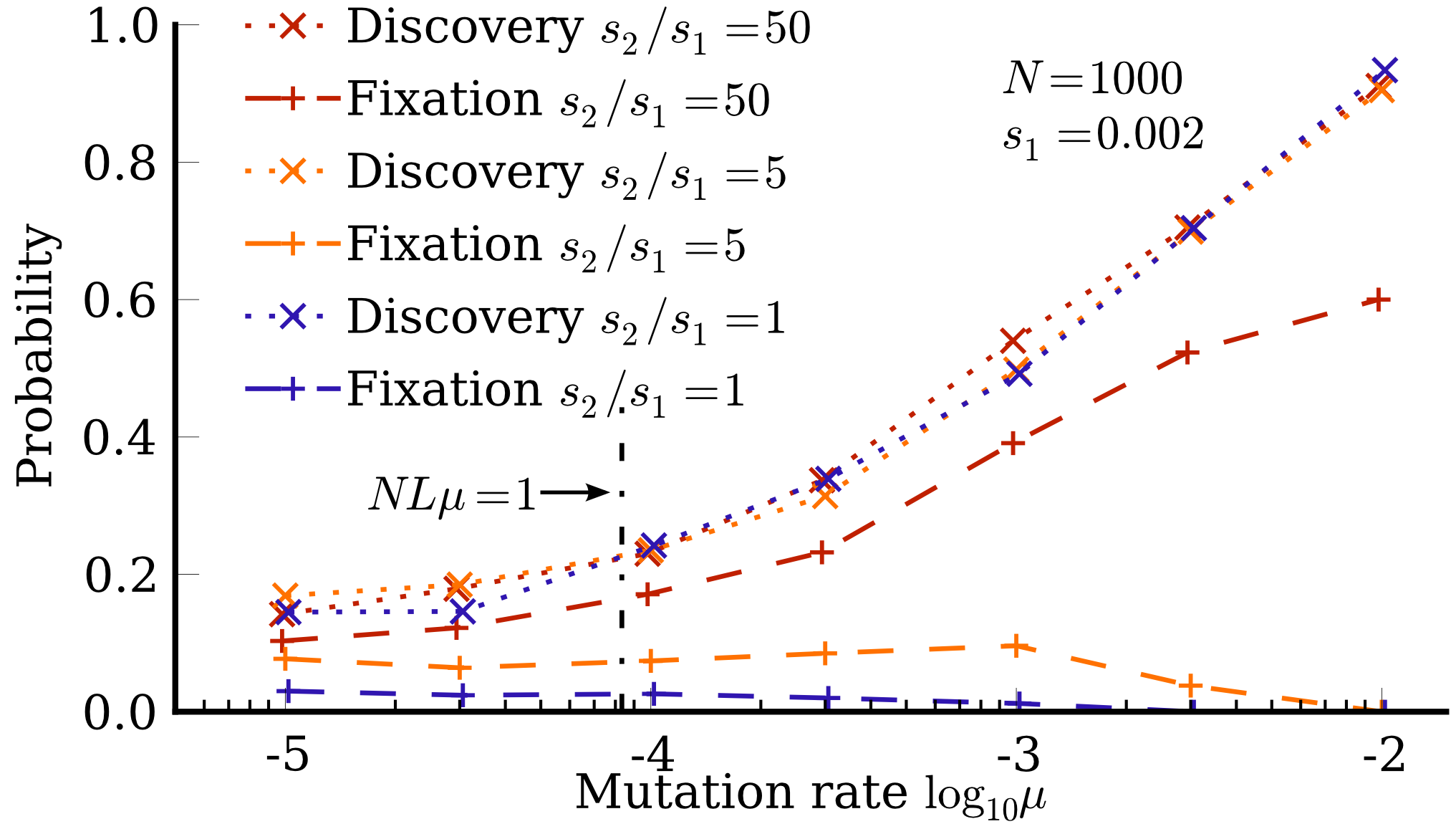} 
\end{center} 
\caption{{\bf The arrival of the frequent.} Probability that phenotype $p2$ is discovered (dotted lines) or is fixed 
(dashd lines) as a function of mutation rate $\mu$ for different relative selection coefficients $s_1/s_2$ for $Ns_1=2$. 
The probability that $p2$ is discovered is independent of relative fitness (within statistical simulation errors). 
Phenotype $p1$ is much more likely to fix than phenotype $p2$, even when the latter is much more fit, due to an ``arrival 
of the frequent'' phenomenon.}
\label{fig:rna_fix} 
\end{figure*}

\subsection*{The arrival of the frequent} 

\par The many orders of magnitude difference in the arrival rate of variation between phenotypes should have many
important implications for evolutionary dynamics. Consider for example the situation where the population has
equilibrated to a phenotype $q$, which was the fitness peak, when subsequently the environment changes so that a
different phenotype $p$ has a higher fitness $1+s$. In order to fix, the alternative phenotype must first be found. If
the time-scale $T_E$ on which the environment changes again is much longer than $T_{p}$ then it likely that the
population will discover and fix $p$. However, if $T_E \ll T_{p}$, then a new phenotype $p'$ may become more fit before
$p$ has time to fix. $T_p$ can vary over many orders of magnitude, so many potentially highly adaptive phenotypes may satisfy $T_p > T_E$  and thus
never be found.

\par Consider also the situation where two phenotypes $p1$ and $p2$ are both more fit than $q$ after an environmental
change. If $s_2 > s_1 \gtrsim 1/2 N$, then in a standard population genetics picture, we would expect $p2$ to fix rather
than $p1$ as long as $T_{p2} \lesssim T_E$. However, this argument ignores the rate at which variation arises. If, for
example, $\phi_{p1q} \gg \phi_{p2q}$, then $p1$ may fix well before $p2$ is discovered and fixes.

\par To illustrate this effect, we study the $L=12$ RNA system depicted in Figure~\ref{fig:RNA}, where the source
neutral space has $N_q=1932$ genotypes, while the two target phenotypes have $\phi_{p1q}=0.067$ and $\phi_{p2q}=0.0015$,
so $\phi_{p2q}/\phi_{p1q} \approx 0.022$, a relatively modest ratio compared to the what could be found from e.g.\
Fig~\ref{fig:mediantime_nullmodel}. For this particular system $\phi_{p1p2} = 0$: there are no direct single mutation
connections between the two target phenotypes -- $p1$ and $p2$ are distinct peaks of the fitness landscape.

\par We simulated a population of $N=1000$ individuals with fixed $s_1= 0.002 > 1/2N$, but with varying ratios $s_2/s_1 \geq 1$. The
population begins on phenotype $q$ and evolves until $p1$ or $p2$ fixed.

Results are shown in Fig.~\ref{fig:rna_fix}. As
the mutation rate increases, and the system moves from the monomorphic to polymorphic regime, the probability that $p2$
is discovered at least once increases (and is largely independent of fitness). Nevertheless, phenotype $p1$ is
discovered much earlier and also much more often because $\phi_{p1q} \gg \phi_{p2q}$. Furthermore, in the monomorphic
regime where $\xi \gg 1$ the population remains on a single genotype $g$ much longer than it takes to explore all the
neighbours. Thus if $p1$ is accessible from $g$, then $p1$ is likely arise repeatedly in relatively quick succession (in ``bursts'').
This effect, which arises naturally in our microscopic model \cite{schaper2013thesis}, can significantly enhance the
probability of fixation over that predicted by origin-fixation models~\cite{yampolsky2001bias} which ignore the
discreteness of the source neutral space.

\par Overall, our simulations show how the more frequent phenotype $p1$ can fix at the expense of the more fit phenotype
$p2$. Given the many orders of magnitude difference possible between the $T_p$, such an ``arrival of the frequent''
effect may prevent the arrival of the fittest: If a highly beneficial phenotype is never discovered, a much less
adaptive but easily accessible phenotype may go to fixation instead.

\par Finally, phenotype $p2$ is significantly less mutationally robust than $p1$ (more frequent phenotypes are typically
more robust~\cite{schaper2011contingency,wagner2008paradox}), and so once discovered, produces deleterious mutants at a
higher rate, making it harder for $p2$ to fix at higher mutations rates, a phenomenon known as ``survival of the
flattest''~\cite{wilke2001flattest}, observed here for the lower ratios $s_2/s_1$ at higher $\mu$. Thus both the
``arrival of the frequent'' and the ``survival of the flattest'' mitigate against the fixation of phenotypes with lower
frequency $F_p$, even if their fitness is much higher.

\par We note that differences in neutral network size have traditionally also been taken into account in terms of free
fitness~\cite{iwasa1988freefitness}, which -- in analogy with free energy in statistical physics
\cite{sella2005statphys} -- incorporates an entropy-like component to account for mutational effects such as genetic
drift and mutational robustness.  This picture provides a theoretical foundation for the ``survival of the flattest''
\cite{wilke2001flattest} effect we observe at high mutation rates in Fig.~\ref{fig:rna_fix}.   However, the ``arrival of
the frequent'' effect is fundamentally different because it  does not rely on mutation-selection balance and
quasi-equilibrium or steady-state assumptions like free-fitness theory does.  Rather, it reflects the strongly
non-equilibrium effect that $p_2$ is rarely or never found.  In the example above, the difference in discovery times between
$p_1$ and $p_2$ is rather modest, and so at large enough mutation rates $p_2$ is found fairly regularly and free-fitness could
be used to analyse results in that regime.   But as can be seen for instance in Fig.~\ref{fig:mediantime_nullmodel} for $L=20$ RNA,
differences in discovery times can vary over many more orders of magnitude than is the case for our particular example, 
so that in practice highly adaptive yet rare
phenotypes may not be discovered at all, even on very long timescales.

\section*{Discussion}

\par Mutations provide the fuel for natural selection. Based on this principle, we have presented a detailed model of
evolutionary dynamics that focuses on a microscopic description of the outcome of mutations. The phenotypic effect of
mutations is mediated by the genotype-phenotype (GP) map which is therefore a crucial ingredient. As outlined in the
introduction, several generic features are shared by many different example maps, independent of model details. Here we
mainly focussed on the fact that these mapping are highly \textit{biased}: Some phenotypes are realised by orders of
magnitude more genotypes than most other phenotypes.

\par Our calculations  for a simplified random mapping and for the more complex RNA secondary structure model predict
that the  large bias observed in the GP maps translates into a similar order of magnitude variation in the median discovery times $T_p$ for a
range of population genetic parameters.  For both maps the local frequencies $\phi_{pq}$ (which predict discovery times) are  a good predictor for the discovery times $T_p$.  For the random GP map $\phi_{pq} \approx F_p$. For RNA this relationship provides a rough first order estimate, but the local frequencies can also deviate strongly, especially when $\phi_{pq}=0$, which can occur even when the global frequency $F_p$ is large.    For both maps a strong bias in the GP map leads 
to a systematic \emph{ordering} of the median discovery times of alternative phenotypes, an effect that we postulate may hold for other GP maps as well.

\par In light of the simplicity of our mean-field approximation, its success in predicting the first-discovery 
time $T_p$ (cf.\ Fig.~\ref{fig:mediantime_nullmodel}) is rather striking. In the random GP map, the excellent agreement
probably arises because all genotypes in the source neutral space are similar in the sense that they
have the same probability distribution to have a certain mutational neighbourhood. There are static fluctuations because the number of neighbours is less than the number of states with $\phi_{pq}\neq 0$.  But while these fluctuations have an effect on processes like fixation, they average out over the many runs used to find the mean or median $T_p$.  By contrast,  in the RNA GP map mutational neighbourhoods of adjacent genotypes are often 
correlated~\cite{huynen1996exploration,wagner2008paradox} so that a single neutral mutation does not completely
re-shuffle the accessible phenotypes (as the mean-field assumption would assume). This effect explains why
the value of the exploration parameter $\gamma$ we obtain by fitting is below the value suggested by our mean-field
model, and also why we still observe around 1 order of magnitude variation in $T_p$ for very similar values of $\phi_{pq}$ (see 
Fig.~\ref{fig:mediantime_nullmodel}). Despite such correlations (which we postulate may occur in other realistic GP maps), rare phenotypes (low $\phi_{pq}$) remain hard to find; 
the strong phenotypic bias in the RNA GP map provides a good a posteriori justification for our mean-field calculations:
The many orders of magnitude range in $\phi_{pq}$ dominates the scale of the phenotype discovery times.

\par The large differences  we observe in the rate with which potential variation appears should have many consequences for
evolutionary dynamics. There is of course a long history of invoking processes that impose directionality on the
pathways available for evolutionary exploration (see ref.~\cite{lynch2007complexity} for a recent discussion). 
Here, by solving microscopic population genetic models, we show in detail just how strong these orienting processes can
be.  Other authors have also pointed out how evolution may favour phenotypes with large neutral networks for RNA, see
e.g.\ refs.~\cite{schuster1994neutralnetworks,cowperthwaite2008ascent}. Similar points have been made for protein models~\cite{ferrada2012comparison}.
 Consider, for example, our $L=20$ RNA system. Despite its rather modest size, we find $10$ orders of magnitude
difference between the discovery times of frequent and rare phenotypes. These differences should
be even more pronounced for larger $L$. In nature, selectable RNA phenotypes are of course characterised by more than
just their secondary structure, and evolutionary processes don't always work at constant $L$. Nevertheless, it is hard
to see how such enormous variations in $T_p$ would not persist in some form in much more sophisticated treatments of
biological RNA. Similar arguments can be made for the other GP maps we listed above. More generally we emphasise that
including the GP map in population genetic calculations may be of importance to a wide range of evolutionary questions.

\par We explicitly showed how phenotypes with a high local frequency can fix at the expense of locally rare phenotypes,
even if the latter have much higher fitness. Taken together, these arguments suggest that the vast majority of possible
phenotypes may never be found, and thus never fix, even though they may globally be the most fit: Evolutionary search is
deeply non-ergodic. When Hugo de Vries was advocating for the importance of mutations in evolution, he famously said
``Natural selection may explain the survival of the fittest, but it cannot explain the arrival of the
fittest''\cite{devries1904mutation}. Here we argue that the fittest may never arrive. Instead evolutionary dynamics can
be dominated by the ``arrival of the frequent".

\section*{Methods}

\par \textit{Simulations} In the dynamic simulations, all $N$ individuals of the population are initially assigned to a
single random genotype in the source neutral space. Then the population evolves for $10N$ generations to reach a steady-state dispersal on the neutral space before measurements are started.

\par \textit{RNA} Secondary structures for RNA  were predicted from sequence using the
Vienna package \cite{ViennaPackage}, version 1.8.5 with all parameters set to their default values.


\bibliographystyle{apsrev}
\bibliography{bibliography}
\newpage

\clearpage

\bigskip

 \section*{Supporting Information}

\bigskip 

\section*{The dependence of $\gamma$ on population dynamic parameters} 
\label{sec:si_fitfactor}

\par For the limiting regimes of polymorphic and monomorphic populations (with $NL\mu \gg 1$ and $NL\mu \ll 1$,
respectively), we have given complete predictions for the first discovery time of an alternative phenotype $p$ depending
on $N$, $\mu$ and $\phi_{pq}$ in Equations~(4) 
 and~(7) 
 Based on these
results, we argued that between these regimes, there should be an interpolating factor $\gamma$ such that a full
expression for the first discovery time is 
\begin{equation} 
	T_p(\alpha) = \frac{-\log(1-\alpha)}{N\gamma L\mu \phi_{pq}}
\label{eq:si_discovery_time_gamma}
\end{equation} 
where all symbols take the same meaning as in the main text.

\par Based on the results in the paper, we predict the following limiting behaviour for $\gamma$: 1) $\gamma \approx 1$
in the large genome limit $L\gg N$; 2) $\gamma \approx 1$ in polymorphic populations ($NL\mu \gg 1$); 3) $\gamma
\rightarrow (K-1)L\rho_q$ in large, monomorphic populations ($NL\mu\ll 1$ but $N\gg L$).

\par Calculating $\gamma$ explicitly is beyond the scope of this work. Instead, we study the behaviour of $\gamma$
numerically through extensive simulations under the random GP map, as outlined above. To this end, we performed 100
simulations for many combinations ($N$,$\mu$) spanning several orders of magnitude for each parameter. The value of
$\gamma$ is calculated from the observed median discovery times $\hat{T}_p$ by a least-squares fit to Equation
\eqref{eq:si_discovery_time_gamma} (with $\alpha=1/2$, as we consider the median discovery times): 
\begin{equation}
\gamma = \frac{\log2 \sum_{p} \phi_{pq} \hat{T}_p}{N L \mu \sum_p (\phi_{pq} \hat{T}_p)^2} 
\end{equation}

\par Figure \ref{fig:si_fitfactors} shows the simulation results, with $\gamma$ multiplied by population size $N$ to
facilitate interpretation. As expected, when $NL\mu \gg 1$, we see that $N\gamma$ approaches $N$, that is $\gamma
\rightarrow 1$: In polymorphic populations, there is little loss of diversity under genetic drift. By contrast, as
$NL\mu$ becomes small, we see that $N\gamma$ tends to $(K-1)L\rho$, showing that in such monomorphic populations, the
localization in genotype space slows down the discovery of alternative phenotypes. Finally, we see that in the large
genome limit ($N=10$, which is smaller than $L=12$), $\gamma$ is roughly independent of $\mu$ and is just below unity,
as we would expect based on our theoretical arguments. The general scaling behaviour of $\gamma$ for intermediate values
of $NL\mu$ is complex and shows no simple dependencies on the dynamic parameter $N$ and $\mu$ (cf. Figures
\ref{fig:si_fitfactors} and \ref{fig:si_fitfactors_unscaled}).

\setcounter{figure}{0} 
\renewcommand{\thefigure}{S\arabic{figure}} 
\renewcommand{\theequation}{S\arabic{equation}}
\renewcommand{\thesection}{S\arabic{section}}

\begin{figure*}
\begin{center}
\includegraphics{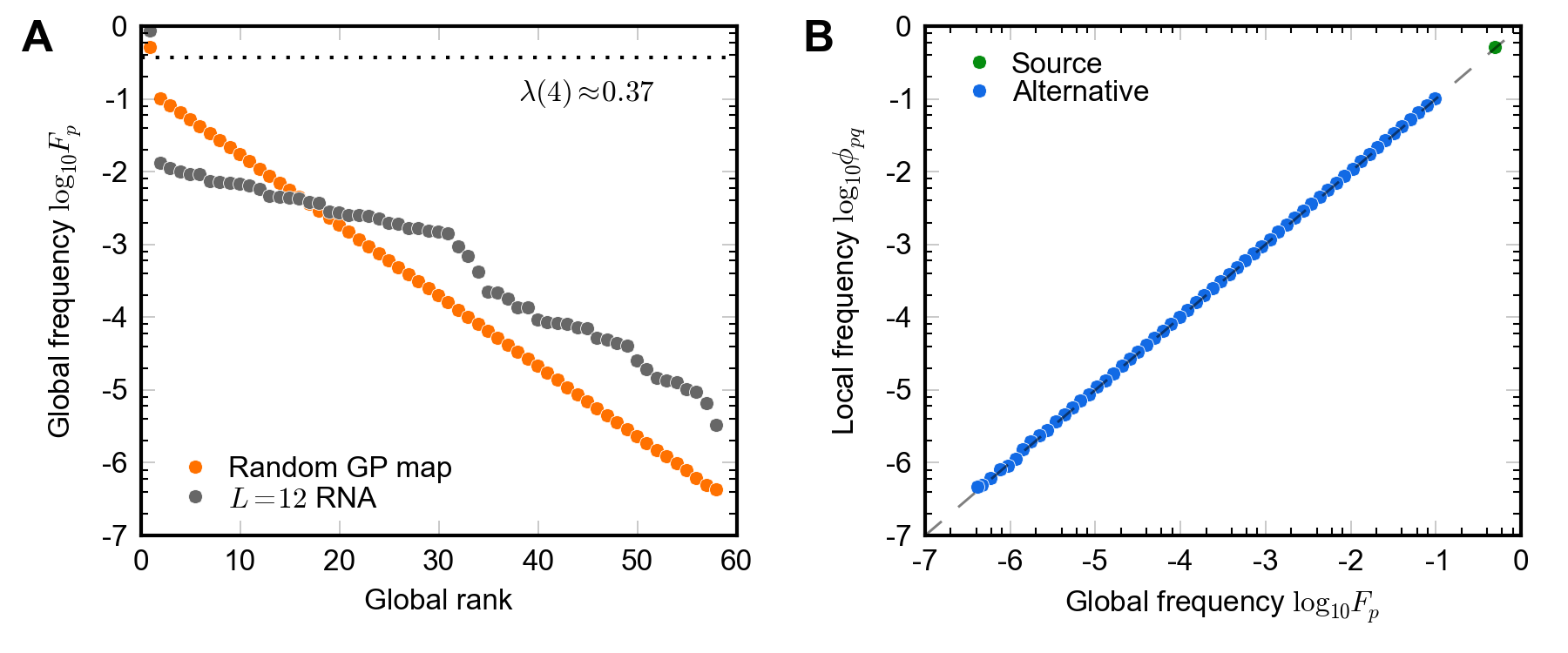}
\end{center} 
\caption{Static properties of the random GP map. \textit{A)} Global
phenotypes frequencies. In addition to the distribution of frequencies $F_p$ used in our simulations (orange), the
diagram also shows the frequencies of RNA secondary structures at $L=12$, obtained by exhaustive enumeration using the
Vienna package, Version 1.8.5 with all parameters set to their default values \cite{ViennaPackage}. \textit{B)}
Comparison of global frequencies $F_p$ and local frequencies $\phi_{pq}$ for the source neutral space $q$ with rank 1.
The robustness of phenotype $q$ ($\rho \equiv \phi_{qq}$) is marked in green; alternative phenotypes ($p\neq q$) are shown
in light blue. The dashed line marks the equality of global and local frequency $F_p = \phi_{pq}$. The relative size of
deviations becomes more severe as $F_p$ becomes small: The less genotypes map into $p$, the less will frozen
fluctuations in the GP map average out.}
\label{fig:si_nullmodel_statics} 
\end{figure*}

\begin{figure*}
\begin{center}
\includegraphics{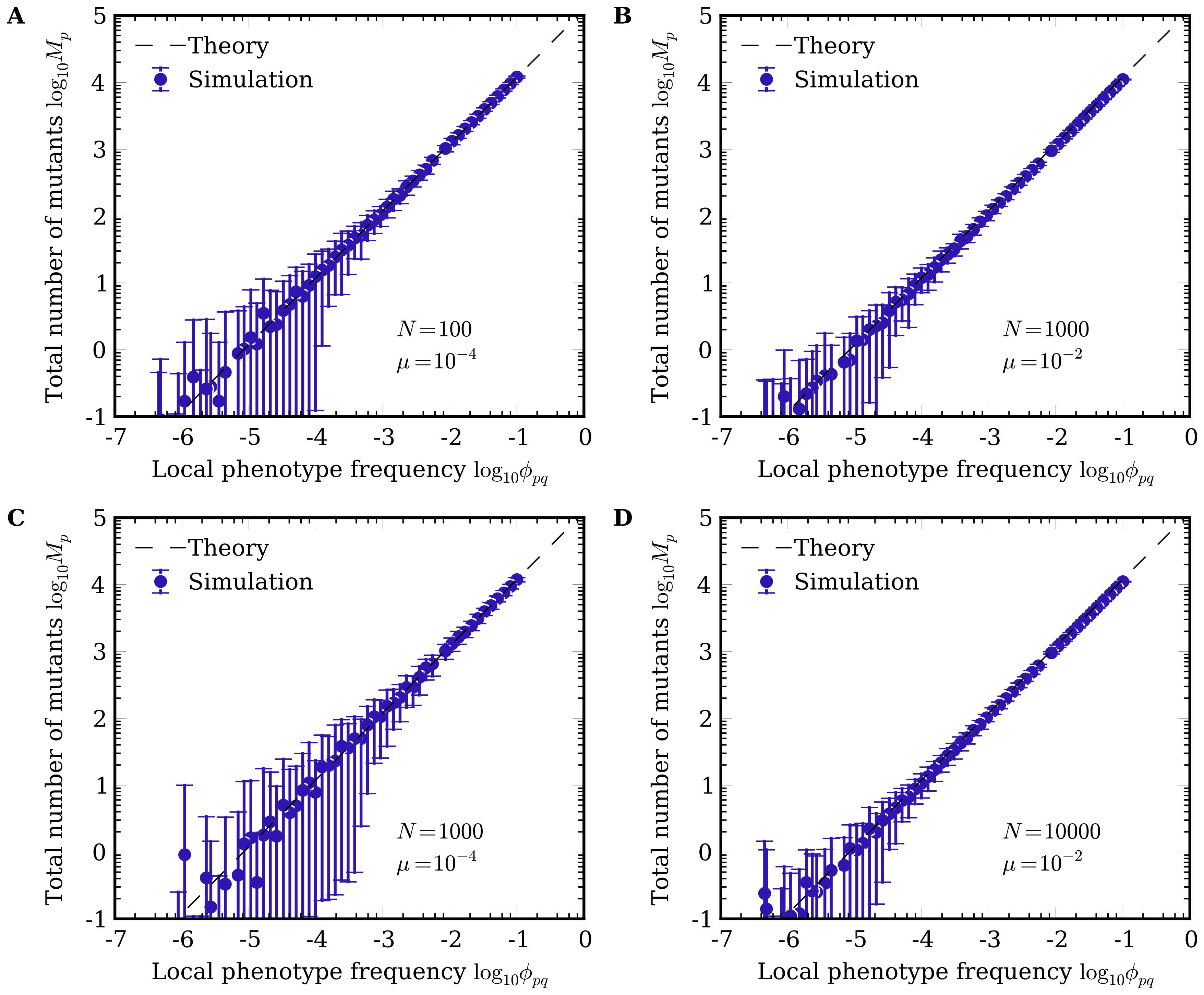}
\end{center} 
\caption{Total number of mutants per phenotype in different dynamic settings. The diagram shows the total
number of mutants $M_p = \sum_{t=1}^T m_p(t)$ carrying phenotype $p$ that were produced during a total of $T=10^4 /
(N\mu)$ generations of simulation under the random GP map. Dots show the average over 100 simulations, error bars show
the standard deviation. The dashed lines correspond to the mean-field theory $M_p=NL\mu\phi_{pq}T$ that follows directly
from Eq.~(3).
In panels B and D, the populations are in the highly polymorphic regime ($NL\mu \gg 1$) and
hence evolve towards greater robustness \cite{vanNimwegen1999robustness} so that the total number of non-neutral mutants
is reduced.} 
\label{fig:si_mutant_count}
\end{figure*}

\begin{figure*}
\begin{center}
\includegraphics{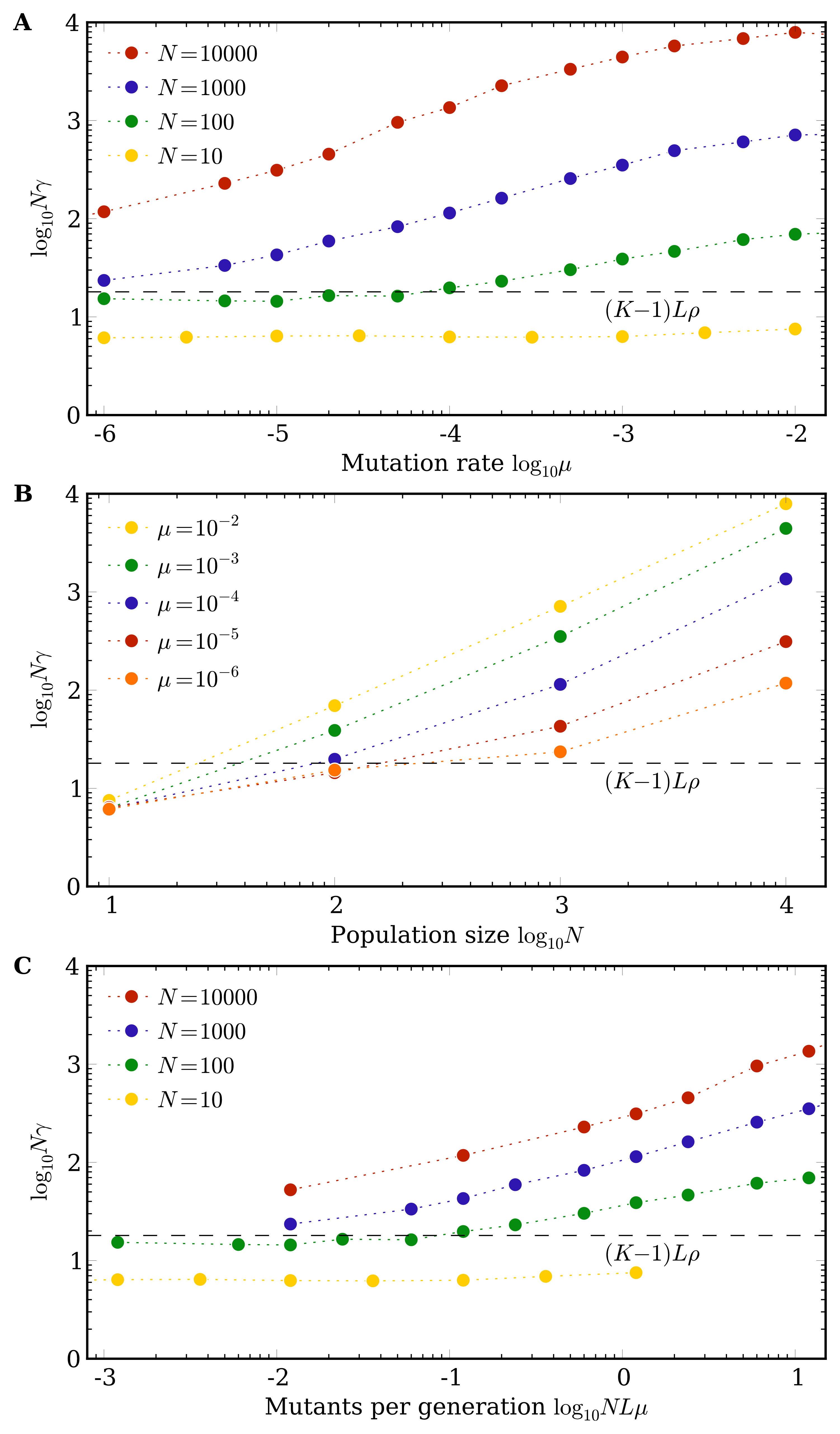}
\end{center} 
\caption{ Scaling of $N\gamma$ with population dynamic parameters. The diagram shows the dependence of
$\gamma$ on: \textit{A}) mutation rate $\mu$, \textit{B}) population size $N$ and \textit{C}) number of mutants per
generation $NL\mu$. Note that the y-axis has been scaled by population size $N$.} 
\label{fig:si_fitfactors}
\end{figure*}

\begin{figure*}
\begin{center}
\includegraphics{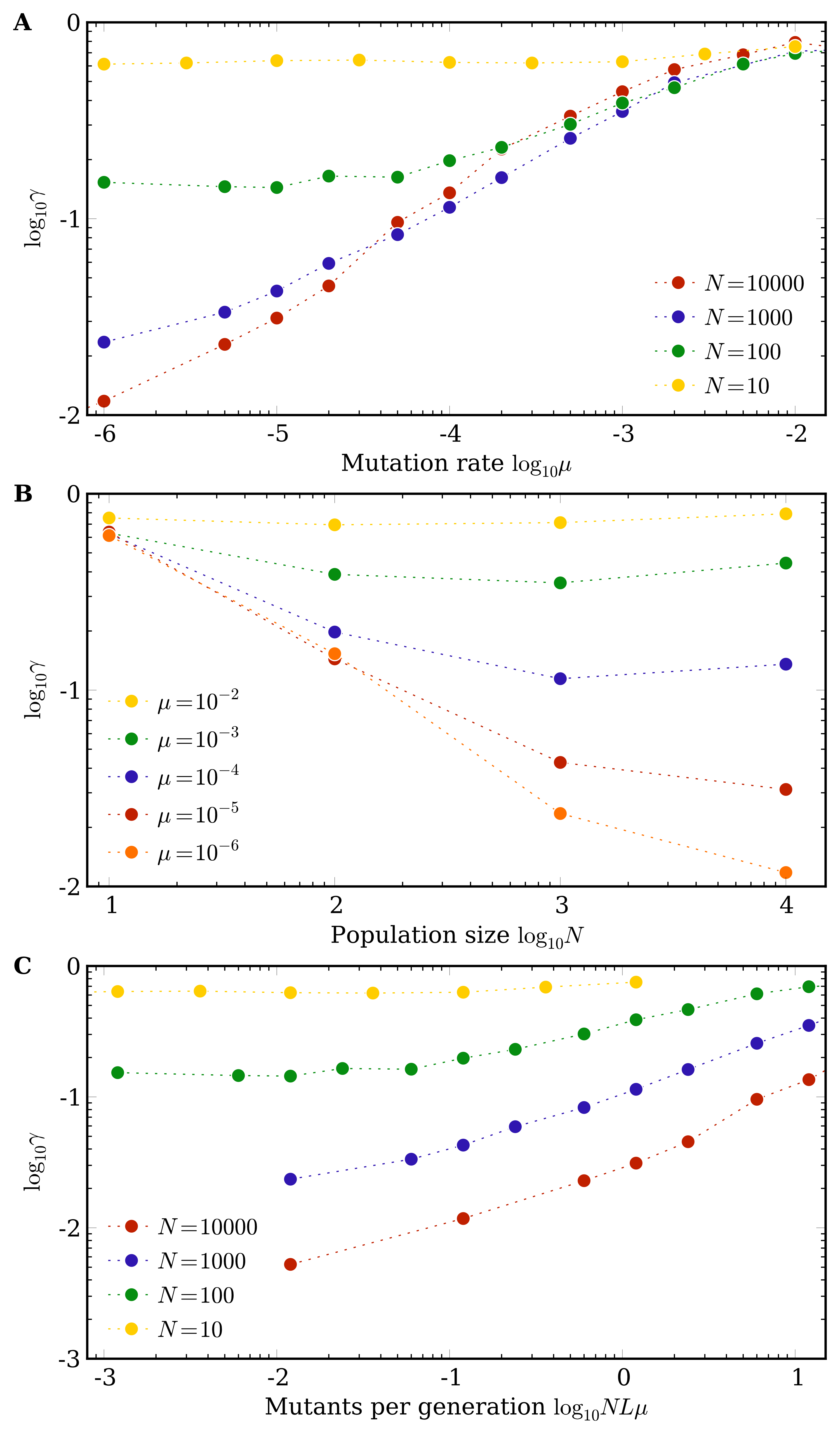}
\end{center} 
\caption{Scaling of $\gamma$ with population dynamic parameters. The diagram shows the dependence of
$\gamma$ on: \textit{A}) mutation rate $\mu$, \textit{B}) population size $N$ and \textit{C}) number of mutants per
generation $NL\mu$. In contrast to Fig. \ref{fig:si_fitfactors}, the y-axis shows $\gamma$ without any scaling factors.}
\label{fig:si_fitfactors_unscaled} 
\end{figure*}

\begin{figure*}
\begin{center}
\includegraphics[height=20cm]{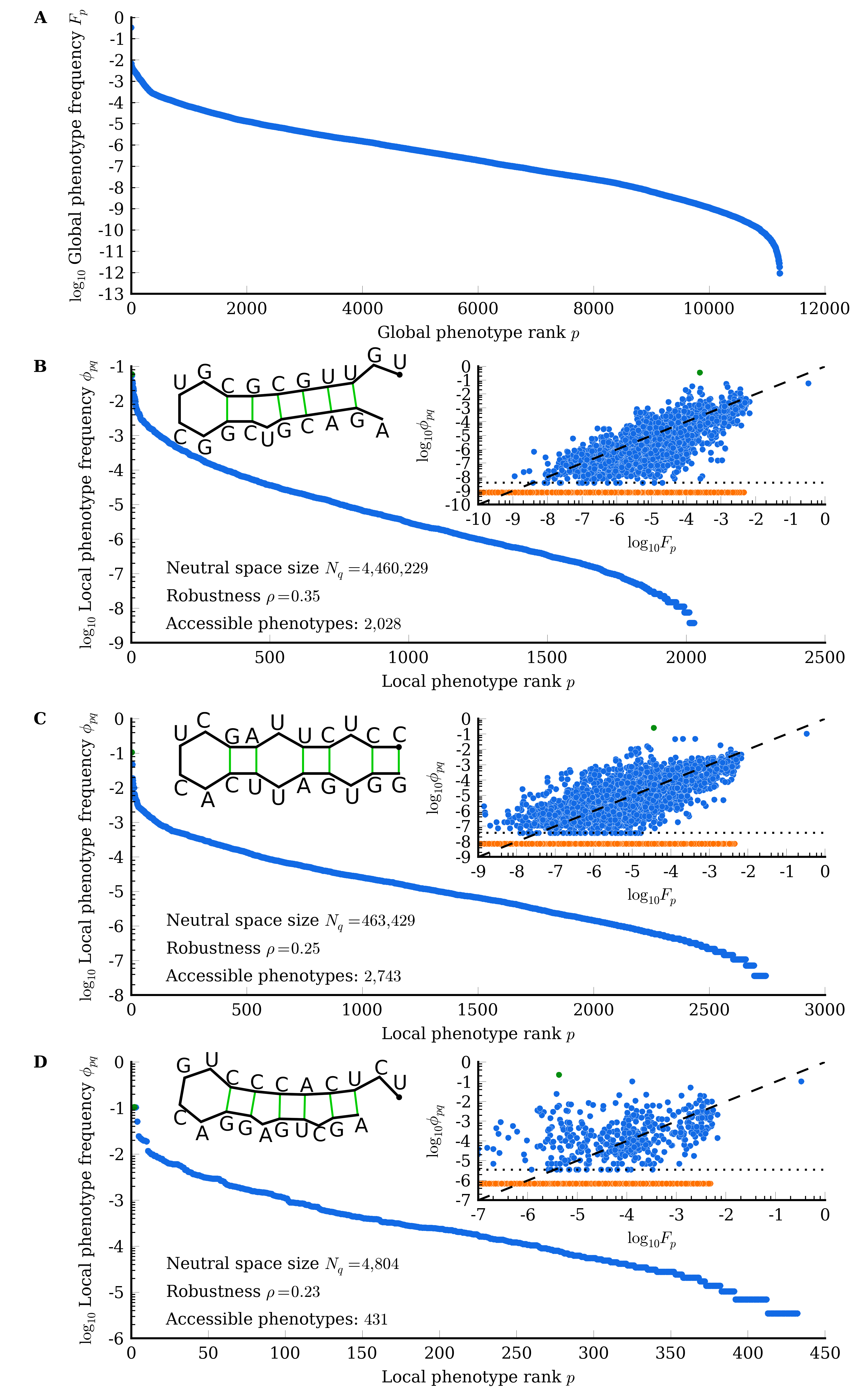}
\end{center}
\caption{Phenotypic bias for RNA secondary structures of length $L=20$. \textit{A}) Global phenotype
frequencies $F_p$ for all $N_P=11,219$ secondary structures. It required about 1 CPU-year on typical present-day
hardware to fold all $4^{20} \approx 10^{12}$ sequences once using the \texttt{fold}-routine of the Vienna package
\cite{ViennaPackage}, version 1.8.5 with all default parameters. \textit{B}-\textit{D}) Local phenotype frequencies
$\phi_{pq}$ around 3 neutral spaces. An example sequence and its secondary structure is given in each panel; starting
from this sequence, the $\phi_{pq}$ can be obtained exactly by tracing out all possible neutral mutations and counting
how often each phenotype is produced. \textit{Insets}: Comparison of global and local frequencies. Accessible phenotypes
($\phi_{pq} > 0$) are drawn in blue, inaccessible phenotypes ($\phi_{pq} = 0$) are shown in orange and the phenotype
corresponding to the neutral space itself is shown in green ($\phi_{qq} \equiv \rho$). The dashed line marks the
equality of local and global frequencies $F_p = \phi_{pq}$ and the dotted line indicates the minimal (non-zero) local
frequency $\phi_{min,q} = 1/(3L N_q)$, corresponding to only a single mutation away from one of the $N_q$ genotypes in
the neutral space. Inaccessible phenotypes with very small global frequencies are omitted for clarity. Note that all these phenotypes are relatively rare ones.}
\label{fig:si_rna_static} 
\end{figure*}

\begin{figure*}[!ht] 
\begin{center}
\includegraphics{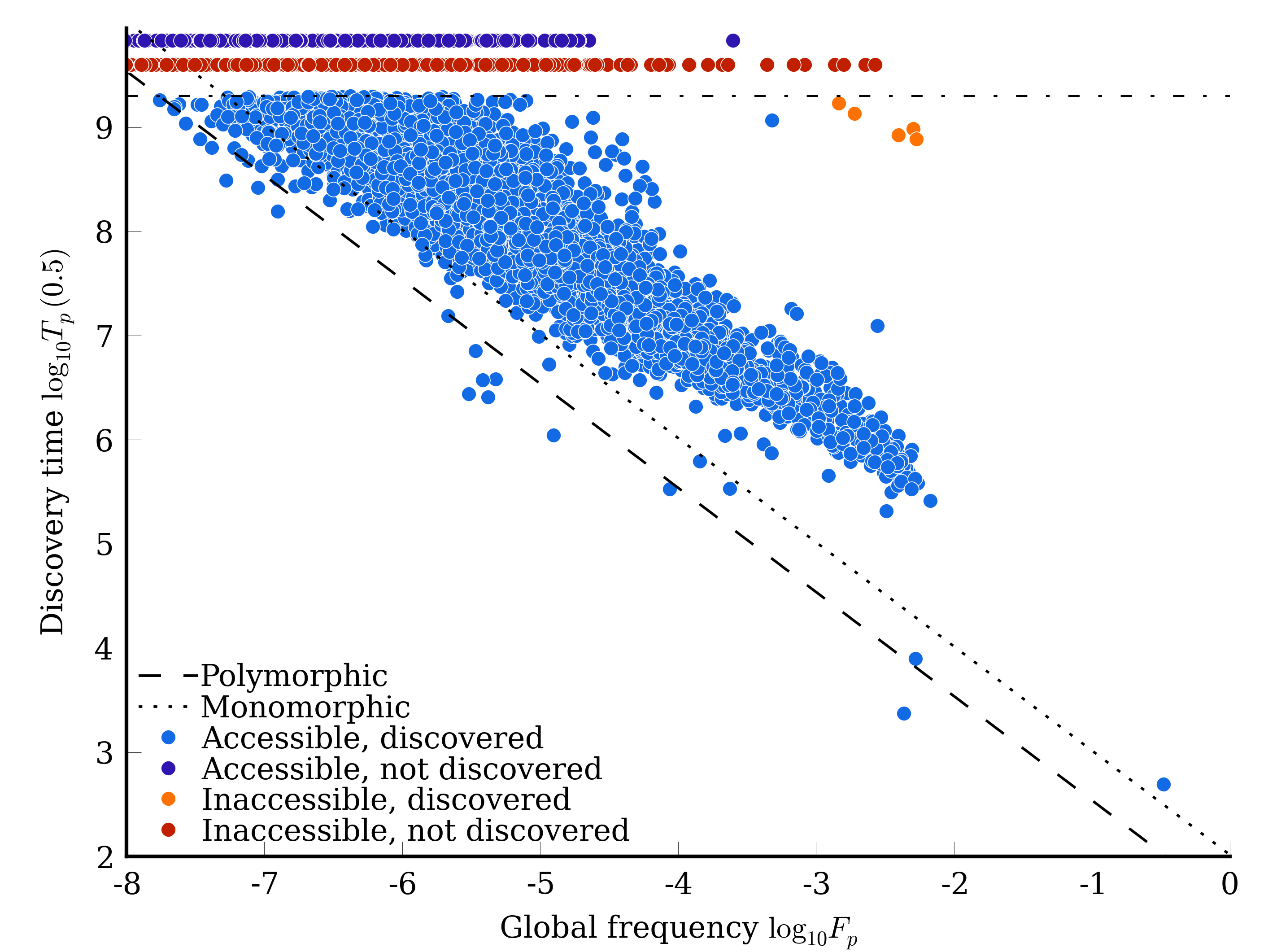}
\end{center} 
\caption{Predictions based on global frequency. The diagram shows the same median discovery times of
alternative RNA secondary structures that are displayed in Fig.~2c,
 but here as a function
of the phenotypes' global frequencies $F_p$ rather than their local frequencies $\phi_{pq}$. The different colors
indicate: Accessible phenotypes that are typically discovered within the simulation time ($T_p(1/2) \leq 2\times 10^9)$,
$\phi_{pq}>0$, light blue); accessible phenotypes that are typically not discovered ($T_p(1/2) > 2\times 10^9$,
$\phi_{pq}>0$, dark blue); inaccessible phenotypes that are typically discovered ($T_p(1/2) \leq 2\times 10^9$,
$\phi_{pq}=0$, orange); inaccessible phenotypes that are typically not discovered ($T_p(1/2) > 2\times 10^9$,
$\phi_{pq}=0$, red). The lines correspond to the prediction for $T_p$ based on global rather than local frequencies:
$T_p(1/2) = \log 2/(NL\mu F_p)$ (cf.\ Eq.~(4)),
dashed) and $T_p(1/2) = \log 2 / (3L^2\mu\rho F_p)$ (cf.\
Eq.~(7)). In contrast to the predictions based on the local frequencies $\phi_{pq}$ in
Fig.~2c,
we note the following: 1) Several phenotypes arise even earlier than predicted by
the analogue of the polymorphic limit (points below dashed line). 2) Many phenotypes are not discovered even though
other phenotypes of comparable (and even much lower) frequency do arise during the simulation. 3) 4 of the most
frequent, but locally inaccessible phenotypes are discovered on a time-scale when double mutations become relevant
(orange dots; since $N=100$ and $\mu=10^{-5}$, double mutants occur on the timescale $t_2 \approx 1/(N(L\mu)^2 =
2.5\times 10^5$, so if double mutations were to lead to globally random phenotypes, we expect phenotypes with $\phi_{pq}
= 0$ to be discovered around $T_p \approx t_2 \log 2/F_p$.)} 
\label{fig:si_rna_TvsF} 
\end{figure*}
\newpage

\end{document}